%% file: main.tex
\documentclass[conference]{IEEEtran}
\usepackage{xcolor}
\usepackage{ifpdf}
\ifpdf
\else
\fi

\usepackage{multirow}

\usepackage{cite}

\renewcommand\IEEEkeywordsname{Keywords}

\ifCLASSOPTIONcompsoc
    \usepackage[caption=false, font=normalsize, labelfont=sf, textfont=sf]{subfig}
\else
\usepackage[caption=false, font=footnotesize]{subfig}
\fi
\ifCLASSINFOpdf
   \usepackage[pdftex]{graphicx}
   \graphicspath{{../pdf/}{../jpeg/}}
   \DeclareGraphicsExtensions{.pdf,.jpeg,.png}
\else
   \usepackage[dvips]{graphicx}
   \graphicspath{{../eps/}}
   \DeclareGraphicsExtensions{.eps}
\fi
\usepackage{amsmath}
\usepackage{array}

\ifCLASSOPTIONcompsoc
  \usepackage[caption=false,font=normalsize,labelfont=sf,textfont=sf]{subfig}
\else
  \usepackage[caption=false,font=footnotesize]{subfig}
\fi
\usepackage{fixltx2e}

\usepackage{stfloats}
\usepackage{url}

\begin{document}
%

\title{Eulerian Phase-based Motion Magnification for High-Fidelity Vital Sign Estimation with Radar in Clinical Settings}



%
\author{\IEEEauthorblockN{Md Farhan Tasnim Oshim\IEEEauthorrefmark{1},
Toral Surti\IEEEauthorrefmark{2},
Charlotte Goldfine\IEEEauthorrefmark{3}, 
Stephanie Carreiro\IEEEauthorrefmark{4},
Deepak Ganesan\IEEEauthorrefmark{1},\\
Suren Jayasuriya\IEEEauthorrefmark{5},and
Tauhidur Rahman\IEEEauthorrefmark{1}}
\IEEEauthorblockA{
\IEEEauthorrefmark{1} Manning College of Information and Computer Sciences, University of Massachusetts Amherst, MA, USA\\
\IEEEauthorrefmark{2}Department of Psychiatry, Yale University School of Medicine, New Haven, CT, USA\\
\IEEEauthorrefmark{3} Department of Emergency Medicine, Brigham and Women’s Hospital, Boston,
MA, USA\\
\IEEEauthorrefmark{4} Department of Emergency Medicine, University of Massachusetts Medical School, Worcester, MA, USA\\
\IEEEauthorrefmark{5} School of Electrical, Computer and Energy Engineering, Arizona State University, Tempe, AZ, USA}}


\maketitle

\begin{abstract}

Efficient and accurate detection of subtle motion generated from small objects in noisy environments, as needed for vital sign monitoring, is challenging, but can be substantially improved with magnification. We developed a complex Gabor filter-based decomposition method to amplify phases at different spatial wavelength levels to magnify motion and extract 1D motion signals for fundamental frequency estimation. The phase-based complex Gabor filter outputs are processed and then used to train machine learning models that predict respiration and heart rate with greater accuracy. We show that our proposed technique performs better than the conventional temporal FFT-based method in clinical settings, such as sleep laboratories and emergency departments, as well for a variety of human postures.
\end{abstract}

\begin{IEEEkeywords}
Motion Magnification; UWB Radar; Gabor Filter, Vital Sign Estimation; Clinical Settings
\end{IEEEkeywords}

%
\IEEEpeerreviewmaketitle

\section{Introduction}

In mobile health (mHealth) research and applications, the detection and magnification of subtle motions are crucial for accurate estimation of physiological parameters. Especially, monitoring of vital signs such as heart rate (HR) involves detection of subtler motion compared to respiration rate (RR). Different sensor modalities can be used to detect and amplify such motions. A number of studies have developed video motion magnification using different optical processing methods including Lagrangian \cite{liu2005motion}, Eulerian linear \cite{wu2012eulerian} , and Eulerian phase-based \cite{wadhwa2013phase} method. However, the use of a camera is obtrusive and can be privacy-invasive when it comes to human motion estimation. The emergence of ultra-wideband (UWB) radar technology gives us the opportunity to unobtrusively monitor motion \cite{javaid2016}, sleep pattern \cite{dopplesleep, sleep-pose-net} as well as vital signs \cite{doc4,farhan2020respiratory} without violating privacy. UWB Impulse Radars perform well on non-contact and continuous monitoring of targets’ macro-movement (translatory motion and limb-level motions) as well as micro-movements (vital signs such as heart rate and respiratory rate). However, these micro-movements become increasingly difficult to detect accurately in the presence of noisy conditions (e.g. in the presence of macro-movements). In order to detect such tiny motions in a contactless manner without any additional equipment, we developed a motion magnification algorithm for radar signals. 

To summarize the key contributions of our paper are as follows: We developed a 1D radar signal magnification pipeline using the Eulerian phase-based magnification algorithm's complex Gabor wavelet pyramid and validated in the following scenarios: (1)  Validation of motion magnification algorithm for synthetic and real radar signals with visualization for known frequency of motion, e.g. pendulum and mass-spring motions (2) Verification of vital sign estimation method in two different settings: sleep laboratory and emergency department.

\begin{figure*}[h]
    \centering
    \includegraphics[width=38pc]{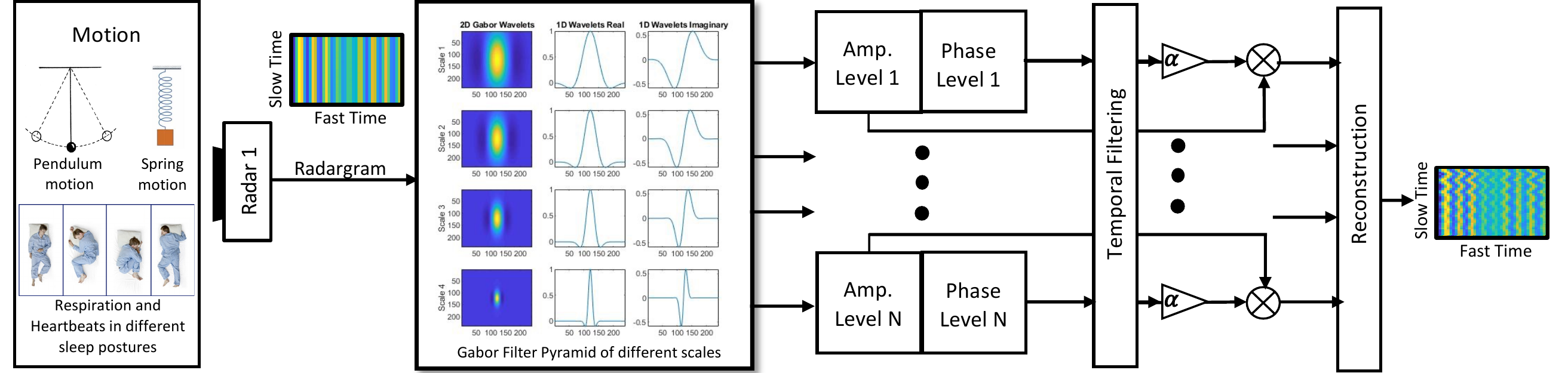}
    \caption{ Phase-based motion extraction pipeline: Captured radargrams are decomposed into bands, temporally filtered, and then added back together to magnify the signal of interest. }
    \label{fig:pipeline}
\end{figure*}

\input{method.tex}

\input{synthetic.tex}

\section{Data Collection Experiments}

To validate our method for respiratory rate (RR) and heart rate (HR) estimation, we collected data in three different settings: a laboratory with actors, an inpatient sleep laboratory with naturally behaving adults, and an emergency department (ED). In the controlled laboratory setting, six adults aging from 23 to 28 years old were asked to lie down on a bed in seven different postures for 1 minute each: 1) lying supine with arms at their sides, 2) supine and browsing on a smartphone, 3) the whole body turned left, 4) the whole body  turned right, 5) fetal position facing left, 6) fetal position facing right, and 7) lying  prone (lying on the stomach). The participants wore a HEXOSKIN wearable body metrics shirt \cite{Hexoskin} for ground truth RR and HR. In the sleep laboratory, two participants underwent complete polysomnography (Siesta, COMPUMEDICS), including electrocardiography and respiratory inductance plethysmography to measure chest and abdominal wall motion. The emergency department data had 14 participants who presented with a variety of conditions requiring conventional cardiopulmonary monitoring. 
Data collection was performed with approvals from the Institutional Review Boards (IRB) of the corresponding institutions.

All settings had multiple (3 each for lab
and ED, and 2 for sleep clinic) PulsON 440 \cite{P440} radars placed at different sides of the bed at least 1m away from the participant. PulsON 440 operated at 3.1-4.8 GHz frequency. We focused on certain range bins of interest where the participant is located (usually in the middle part of the bed). The phase-based algorithm is applied to extract 14 1D signals (one for each wavelength of the Gabor filter) of the length of the window size. Then FFT peak and zero-crossing rate for each 1D time series were calculated and a total of 28 features were used to train the machine learning algorithms. We selected the best features for RR and HR prediction with different regression algorithms. For each of the vital sign estimations, a windowing of 30 seconds with a window shift of 5 seconds were employed.

\section{Results}
\begin{figure}
    \centering
    \includegraphics[width=0.48\textwidth]{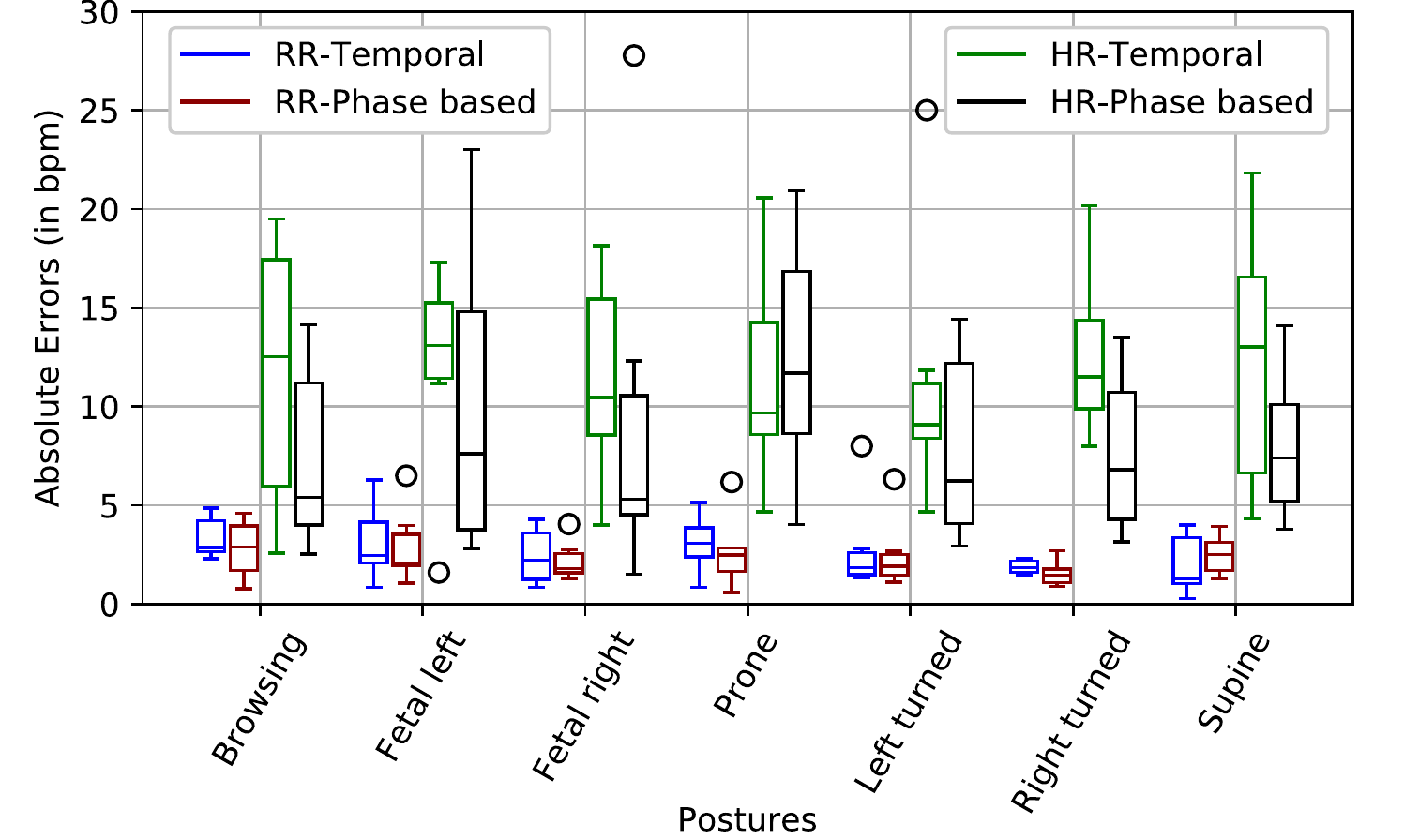}
    \caption{Phase-based RR and HR performance compared to Temporal method over different sleep postures}
    \label{fig:BR_HR_box}
\end{figure}

For obtaining the best results, the frequency corresponding to the highest FFT peak of filtered 1D signal was used in addition to the zero-crossing rates with Gabor filters having spatial wavelength of 75, 15, 10, 9, 7, 5, and 4.
In all cases, the spatial bandwidth of one-fifteenth of the wavelength was employed. The performance of the proposed phase-based algorithm with the temporal FFT-based baseline method for both respiratory rate and HR estimation is shown in Figure \ref{fig:BR_HR_box}. The box plots show the absolute error in the estimated RR and HR across different sleeping postures as prior work has identified posture to be an important factor influencing the performance \cite{dopplesleep}. While the phase-based technique performs slightly better for RR estimation in all postures except supine, the proposed phase-based technique yields a  significantly higher performance increase for subtler motions like heartbeats as shown in Table \ref{tab:results}. The HR prediction using the Gabor features and logistic regression performed better than the conventional temporal technique by at least 4.66 bpm lower MAE in the lab and at least 3.7 bpm lower MAE in sleep clinic. The improvement in HR estimation with phase-based ML technique compared to temporal technique is also reflected across every sleeping posture as illustrated in Figure \ref{fig:BR_HR_box}.  


 


\begin{table}[]
\centering
\caption{RR and HR Performance of different regression models using 10-fold cross-validation in comparison with baseline temporal FFT model on three different environment settings}
\begin{tabular}{|l|ccc|ccc|}
\hline
\multicolumn{1}{|c|}{\multirow{2}{*}{Method}} & \multicolumn{3}{c|}{\begin{tabular}[c]{@{}c@{}}RR\\ MAE in bpm\end{tabular}} & \multicolumn{3}{c|}{\begin{tabular}[c]{@{}c@{}}HR\\ MAE in bpm\end{tabular}} \\ \cline{2-7} 
\multicolumn{1}{|c|}{}                        & Lab                      & Sleep                     & ED                   & Lab                         & \multicolumn{2}{c|}{Sleep}                     \\ \hline
Temporal FFT                                    & 2.65                     & 2.99                      & 4.51                     & 11.65                       & \multicolumn{2}{c|}{7.98}                      \\
RF with Gabor features                        & 1.90                     & 1.48                      &        3.69               & 7.42                        & \multicolumn{2}{c|}{4.28}                      \\
LR with Gabor features                        & 2.07                     & 2.02                      &      4.23                 & 6.99                        & \multicolumn{2}{c|}{4.28}                      \\ \hline
\end{tabular}
\\
\vspace{5pt}
RF: Random Forest, LR: Linear Regression, ED: Emergency Department
\label{tab:results}
\end{table}

\section{Conclusion and Future Work}


We developed a motion extraction pipeline to selectively amplify subtle motion in a background of larger motion using Gabor filters of different spatial wavelengths. Additionally, to improve vital sign estimation over conventional temporal techniques, we incorporated machine learning to the FFT peaks and zero crossing features extracted from phase-based Gabor filters. Future advances in vital sign estimation could be achieved by training a deep learning algorithm with the complex-valued Gabor filtered and magnified signals. 

\section*{Acknowledgment}
This work is in part supported by the National Science Foundation under grant SBE 1839999, grant SCH 2124282, TSA grant UL1 TR001863 from the National Center for Advancing Translational Science (NCATS, components of the National Institute of Health), NARSAD Brain and Behavior Research Foundation (TS), and start-up grant support from the Manning College of Information and Computer Sciences and the Institute for Applied Life Sciences at the University of Massachusetts Amherst.




\bibliographystyle{IEEEtran}
\bibliography{ref}
%



\end{document}

%% file: method.tex
\section{Phase-based Motion Extraction Pipeline}

Our method is primarily inspired by phase-based video motion magnification introduced by Wadhwa et al.~\cite{wadhwa2013phase} that uses complex-valued steerable pyramids~\cite{simoncelli1992shiftable,portilla2000parametric} to spatially decompose a video into amplitude and phase components. Phase at each level of the pyramid corresponds to local motion, and then are temporally bandpass filtered, amplified, and added back onto itself. 
The major challenge with adapting phase-based motion extraction to UWB is the lack of spatial data necessary to construct a complex-valued steerable pyramid. Instead, we require a 1D version of the phase-based motion magnification. Following~\cite{wadhwa2013phase}, we explicate this 1D motion magnification for a simple translational model and a periodic signal with a valid Fourier series decomposition. Given original signal as $f(x)$ at time $t=0$, the small motion can be represented as $f(x+\delta(t))$, a small translation in space. Thus, to amplify motion, we seek to compute $f(x + (1+\alpha)\delta(t))$ corresponding to the magnified translational movement. The Fourier series decomposition of $f$ yields the following~\cite{wadhwa2013phase}:
\begin{equation}
    f(x+\delta(t)) = \sum_{\omega = -\infty}^{\infty} A_{\omega}e^{i\omega(x+\delta(t))}
\end{equation}
for coefficients $A_{\omega}.$ Note that for a particular frequency band $\omega$, we can extract the phase of that band as $\omega\delta(t)$, and thus multiplying our signal by $e^{i\alpha \omega \delta(t)}$ yields the following Fourier series:
\begin{equation}
    \sum_{\omega = -\infty}^{\infty} A_{\omega}e^{i\omega(x+\delta(t))} e^{i\alpha \omega \delta(t)} = f(x+(1+\alpha)\delta(t))
\end{equation}

In this paper, we refer to the above method as global phase-based motion magnification. This can be accomplished for non-periodic signals using a Fast Fourier Transform (FFT), extracting the phase, temporally bandpass filtering the phase, and amplifying back onto itself before computing the inverse FFT. 
Just as video motion magnification requires an image pyramid to analyze local motion, we require amplifying small motions in our 1D radar signals. To do so, we utilize 1D Gabor wavelets to analyze phases locally allowing us to build the analog of a complex-valued steerable pyramid for radar signals. 
Our task of subtle motion magnification from radar reflections is inherently 1D in nature and hence does not require the steerability property of the filter. Thus we use 1D Gabor filter decomposition for our phase-based processing. The 1D Gabor filter can be defined by the following equation:
\begin{equation}
\begin{aligned}
 g(x)&=\frac{1}{\sqrt{2\pi}\sigma^2}e^{-\frac{x^2}{2\sigma^2}}e^{i\omega x} 
 \end{aligned}
 \end{equation}
 
To obtain a pyramid representation, we need to compute these Gabor wavelets at different spatial scales. This is done primarily by varying $\sigma$ (typically in powers of 2), for the number of layers desired in the pyramid. The resulting Gabor wavelets are shown in Figure~\ref{fig:pipeline}. 

To reconstruct the pyramid, we utilize the property of steerable pyramids~\cite{wadhwa2013phase} that $f(x) = \sum \hat{f}(\omega)\Psi_{\omega}^2,$ where $\hat{f}$ is the discrete Fourier transform of the signal, and $\Psi_\omega$ is the Fourier transformed Gabor wavelet. Since $\hat{f}\Psi$ corresponds to the convolution of the Gabor wavelet with the original signal in the primal domain (via the Fourier convolution theorem), $\hat{f}\Psi^2$ corresponds to convolving the result again with Gabor wavelets to reconstruct the original signal. This is a useful property of Gabor wavelets, which is computationally efficient and elegant to perform the reconstruction. 
 
Thus our full processing pipeline for magnification outlined in Figure~\ref{fig:pipeline} is as follows: (1) magnify each radar signal along the slow-time dimension (each column of the radargram) by convolving the signal with multiple Gabor wavelets of different scales, (2) compute the amplitude and phase at each level of the pyramid, (3) bandpass filter the resulting phase signals for the desired motion frequency, (4) amplify the phase and multiply it to the amplitude component, and (5) reconstruct the signal by convolving each level with the same Gabor wavelet and adding the levels back up after aligning them in phase to avoid destructive interference.


%% file: synthetic.tex
\section{Data Analysis with Motion Magnification}
\label{sec:synthetic}

\begin{figure}[htb!]
\centering
    \includegraphics[width=0.46\textwidth]{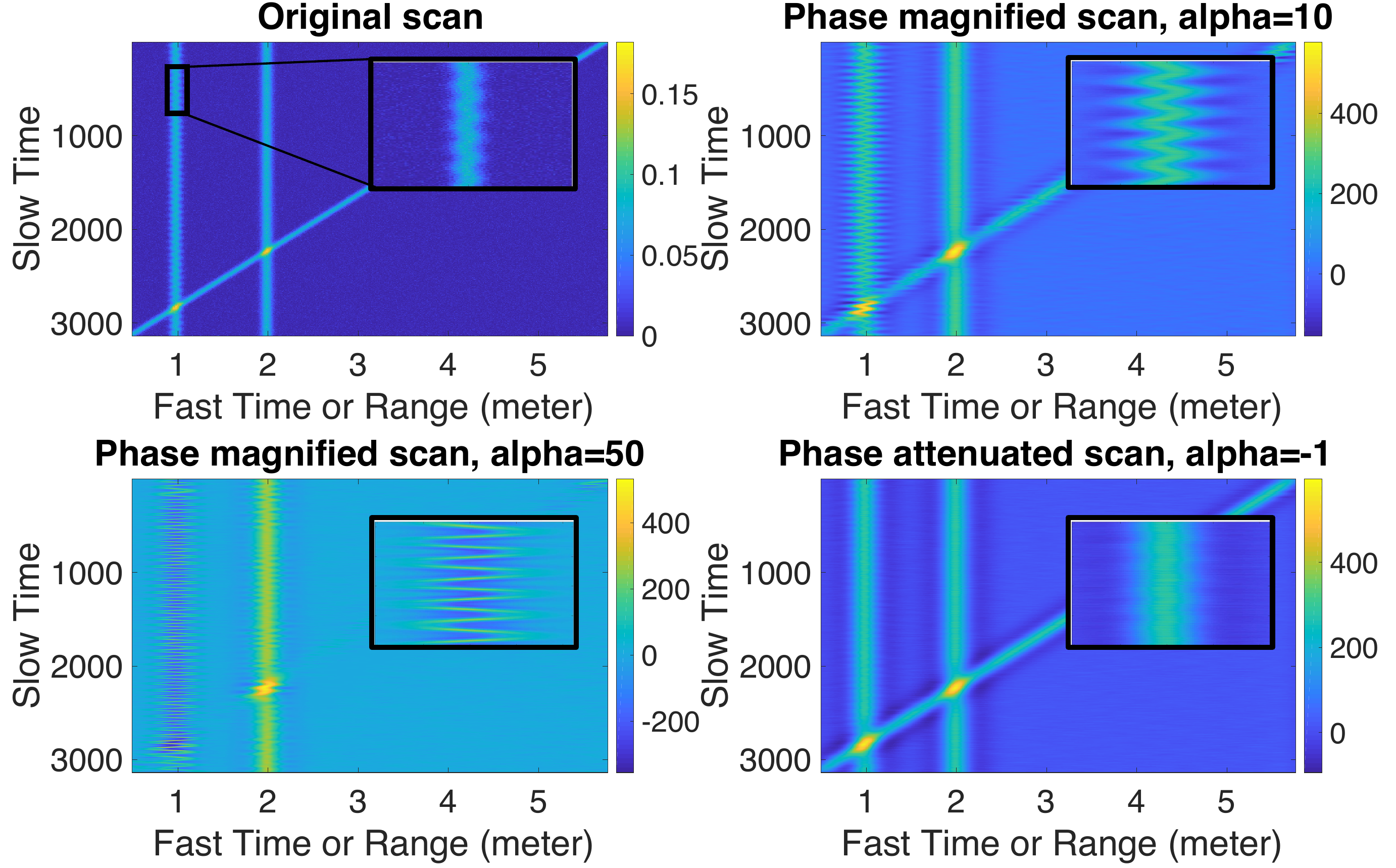}
    \caption{Synthetic data analysis with Eulerian phase-based technique}
     \label{fig:phasesim}
\end{figure}


To test the effectiveness of our method, we perform synthetic data analysis using simulated radar measurements. 
We modeled a target moving with a tiny sinusoidal motion with a frequency 45Hz having amplitude from 1 range bin to a hundredth of a range bin centered at a 1m distance from the radar. We also created two non-sinusoidal motions acting as constant background reflectors 2m away from the radar as well as a diagonal motion proceeding towards the radar at constant velocity. Our method aims to selectively magnify the motion of the moving target without affecting the non-targets: constant reflector's signal and diagonal motion of constant velocity. In Figure~\ref{fig:phasesim}, we show the radargrams of our original signal on the top left along with magnified versions at amplification factors $\alpha = 10$ and $\alpha = 50$ with our method on the top right and bottom left respectively. Note how the displacement of the target increases in the inlet as $\alpha$ increases. We also perform motion attenuation by choosing $\alpha < 0$, thus dampening motion at that frequency in the scene as shown in the bottom right of Figure~\ref{fig:phasesim}. Therefore, we show by simulation that our method is capable of magnifying the subtle sinusoidal motion without magnifying the non-sinusoidal motions.